

\magnification=1200
\hsize=13.5cm

\def\sk{\vskip .2cm}
\def\noi{\noindent}

\def\al{\alpha}

\def\unmezzo{{1 \over 2}}
\def\epsi{\varepsilon}
\def\we{\wedge}

\def\de{\delta}

\def\part{\partial}

\def\Lcal{{\cal L}}

\def\R#1#2{ \Lambda^{#1}_{~~#2} }
\def\Rinv#1#2{ (\Lambda^{-1})^{#1}_{~~#2} }
\def\PS#1#2{ {P_S}^{#1}_{~~#2} }
\def\PA#1#2{ {P_A}^{#1}_{~~#2} }

\def\C#1#2{ {\bf C}_{#1}^{~~~#2} }
\def\c#1#2{ C_{#1}^{~~#2} }
\def\q#1{   {{q^{#1} - q^{-#1}} \over {q^{\unmezzo}-q^{-\unmezzo}}}  }

\def\Cb{{\bf C}}
\def\q1{$q \rightarrow 1$}
\def\Fmn{F_{\mu\nu}}
\def\Am{A_{\mu}}
\def\Ama{A_{[\mu}}
\def\An{A_{\nu}}
\def\dm{\part_{\mu}}
\def\dn{\part_{\nu}}
\def\Ana{A_{\nu]}}

\def\dma{\part_{[\mu}}

\def\su{$SU(2) \times U(1)~$}
\def\gij{g_{ij}}
\def\Lcal{{\cal L}}

\def\alfa{ {2\over {2+q^2+q^{-2}}}}
\def\La{\Lambda}
\def\Lam{\La^{-1}}

\hskip 10cm \vbox{\hbox{DFTT-74/92}\hbox
{December 1992}}
\vskip 0.6cm
\centerline{\bf  $U_q(N)$ GAUGE THEORIES }
\vskip 3cm
\centerline{\bf Leonardo Castellani}
\vskip .4cm
\centerline{\sl Istituto Nazionale di Fisica Nucleare}
\centerline{\sl and}
\centerline{\sl Dipartimento di Fisica Teorica}
\centerline{\sl Via P. Giuria 1, I-10125 Torino, Italy}
\vskip 2cm
\centerline{\bf Abstract}
\sk
Improving on an earlier proposal, we construct the gauge theories of the
quantum groups $U_q(N)$. We find that these theories are consistent also
with an ordinary (commuting) spacetime. The bicovariance conditions of
the quantum differential calculus are essential in our construction.
The gauge potentials and the field strengths are
$q$-commuting ``fields", and satisfy
$q$-commutation relations with the gauge parameters.
The transformation rules of the potentials are given explicitly,
and generalize the ordinary infinitesimal gauge
variations. The $q$-lagrangian
invariant under the $U_q(N)$ variations
has the Yang-Mills form $\Fmn^i \Fmn^j
g_{ij}$,  the ``quantum metric'' $g_{ij}$ being a
generalization of the Killing
metric.

\vskip 3.3cm
\vbox{\hbox{DFTT-74/92}\hbox{December 1992}}
\sk
\hrule
\vskip .2cm
\leftline{{\sl e-mail addresses:} 31890::castellani,
castellani@torino.infn.it}
\vfill
\eject

In recent times there have been some attempts [1] to
construct gauge theories of quantum groups [2]. These theories
are continuously
connected with ordinary Yang-Mills theories, just as quantum groups are
continuously connected with ordinary Lie groups.
In ref. [3] we have proposed a geometric approach to the construction of
$q$-gauge theories, based on the differential calculus on
$q$-groups developed in
[4-9], and we have found a $q$-lagrangian
invariant under the action of $[SU(2) \times U(1)]_q$. Here we
generalize the procedure to any quantum
group $U_q(N)$.
\sk
Spacetime is taken to be ordinary
(commutative) spacetime, but the whole discussion holds for
a generic $q$-spacetime.
\sk
We start by recalling some essential
facts about quantum Lie algebras. There are
many ways to deform Lie algebras: here
we consider the one that naturally comes
from the quantum differential
calculus, i.e. the $q$-algebra of the quantum tangent
vectors. The notations will be as in ref. [9].
\sk
The $q$-algebra we obtain from a
bicovariant differential calculus is given
by the $q$-commutations between the quantum generators $T_i$:

$$T_i T_j - \R{kl}{ij} T_k T_l = \C{ij}{k} T_k \eqno(1) $$

\noi The braiding matrix $\Lambda$  and the $q$-structure
constants $\C{ij}{k}$
can be expressed in terms
of the $R$-matrix of the corresponding $q$-group, as
shown in [6] and further discussed in [7-9], and satisfy four conditions:

$$\R{ij}{kl} \R{lm}{sp} \R{ks}{qu}=\R{jm}{kl} \R{ik}{qs} \R{sl}{up}
{}~~~~~~~~{\rm(Yang-Baxter~equation)} \eqno(2a)$$
$$\C{mi}{r} \C{rj}{n} - \R{kl}{ij} \C{mk}{r}
\C{rl}{n} = \C{ij}{k} \C{mk}{n}
{}~~~{\rm (q-Jacobi~identities)} \eqno(2b) $$
$$\C{is}{j} \R{sq}{rl} \R{ir}{pk} + \C{rl}{q} \R{jr}{pk} = \R{jq}{ri}
\R{si}{kl} \C{ps}{r} + \R{jq}{pi} \C{kl}{i} \eqno(2c)$$
$$\R{ir}{mk} \R{ks}{nl} \C{rs}{j}=\R{ij}{kl} \C{mn}{k} \eqno(2d)$$

\noi  cf. for ex. [4,5,9]. The last
two conditions are trivial in the limit \q1 ($\R{ij}{kl}
=\de^i_l \de^j_k$). The q-Jacobi identity shows that the matrix
$(T_i)_j^{~k} \equiv \C{ji}{k}$ is a representation (the adjoint
representation) of the q-algebra (1).
\sk
This is all the information we need
from the general theory of [4]. We introduce
now the definition of field strength

$$\Fmn \equiv \unmezzo (\dm \An - \dn \Am + \Am \An - \An \Am)=
        \dma \Ana - \Ama \Ana \eqno(3)$$

\noi where

$$\Am \equiv \Am^i T_i$$

\noi The gauge potentials $\An^i$ satisfy the $q$-commutations:

$$A^i_{[\mu} A^j_{\nu]} =
-{1\over{q^2+q^{-2}}} (\La+\La^{-1})^{ij}_{~~kl}
A^k_{[\mu} A^l_{\nu]}\eqno(4)$$

\noi and simply commute with the quantum
generators $T_i$. The square parentheses
around the $\mu$, $\nu$ indices stand
for ordinary antisymmetrization. The inverse
$\Lam$ of the braiding matrix always exists [4] and is defined
by $\Rinv{ij}{kl} \R{kl}{mn}=\de^i_m \de^j_n$.
The rule (4) is
inspired by the $q$-commutations of
exterior products of left-invariant one-forms
on the quantum groups $U_q(N)$:

$$A^i \we A^j = -{1\over{q^2+q^{-2}}} (\La+\La^{-1})^{ij}_{~~kl}
A^k \we A^l\eqno(5)$$

\noi deduced in ref. [9]. Then replacing

$$A^i=A^i_{\mu} dx^{\mu} \eqno(6)$$

\noi into eq. (5) yields the
rule (4). Note that the differentials $dx^{\mu}$ are
ordinary spacetime differentials,  commuting with $A^i_{\mu}$.
\sk
For future use we define the projectors:

$$\PS{ij}{kl} \equiv {1\over {2+q^2+q^{-2}}}
[(q^2+q^{-2})+(\La+\La^{-1})]^{ij}_{~~kl}
    \eqno(7a)$$
$$\PA{ij}{kl} \equiv {1\over {2+q^2+q^{-2}}}
[2I-(\La+\La^{-1})]^{ij}_{~~kl}
    \eqno(7b)$$

\noi where $I^{ij}_{~~kl} \equiv
\de^i_k \de^j_l$ is the identity matrix.
It is not difficult to verify the properties:

$$P_S P_S=P_A P_A=I,~~~P_S P_A=P_A P_S=0\eqno(8)$$

 \noi due to the spectral decomposition of $\La$ [7]:

 $$(\La+q^2 I)(\La +q^{-2}I)(\La-I)=0, \eqno(9)$$

 \noi a consequence of the Hecke relation for the $U_q(N)$ $R$-matrix
[2c].
 \sk
We are now ready to compute the explicit form of the field strength:
$$\eqalign{
\Fmn^i T_i=&\dma \Ana^i T_i + \Ama^i \Ana^j T_i T_j =\cr
=&\dma \Ana^i T_i + \al \Ama^i \Ana^j T_i T_j + (
1-\al) \Ama^i\Ana^j T_iT_j=\cr
=&\dma \Ana^i T_i + \al \Ama^i \Ana^j T_i T_j -
{(1-\al) \over {q^2+q^{-2}}}
(\La+\La^{-1})^{kl}_{~~ij}\Ama^i\Ana^j T_kT_l \cr
=&\dma \Ana^i T_i + \al\Ama^i \Ana^j
[T_i T_j - {(1-\al) \over {\al(q^2+q^{-2})}}
(\La+\La^{-1})^{kl}_{~~ij} T_kT_l] \cr}\eqno(10)$$
\noi where we have used the
$q$-commutations (4). Setting $\al=\alfa$ we find:

$$\eqalign{
 \Fmn^i T_i
 =&\dma \Ana^i T_i + \alfa\Ama^i \Ana^j  [T_i T_j -\unmezzo
(\La+\La^{-1})^{kl}_{~~ij} T_kT_l] \cr
=&\dma \Ana^i T_i +
{1\over{2+q^2+q^{-2}}}\Ama^i \Ana^j  [\C{ij}{n}-\Rinv{kl}{ij}
\C{kl}{n}] T_n \cr} \eqno(11)$$

\noi since

$$T_i T_j - \unmezzo (\La + \Lam)^{kl}_{~~ij}T_k T_l = \unmezzo
[\C{ij}{n}-\Rinv{kl}{ij} \C{kl}{n}] T_n\eqno(12) $$

\noi (use (1) and the relation one obtains after multiplying
(1) by $\La^{-1}$). Let us define

$$ \C{kl}{n} \equiv \c{kl}{n} - \R{ij}{kl} \c{ij}{n} \eqno(13)$$

\noi Then

$$\C{ij}{n}-\Rinv{kl}{ij} \C{kl}{n} =
(2+q^2+q^{-2}) \PA{kl}{ij} \c{kl}{n}
\eqno(14)$$

\noi and the field strength can finally be written as:

$$\Fmn^i =\dma \Ana^i + \PA{kl}{mn} \c{kl}{i} \Ama^m \Ana^n  \eqno(15)$$

Note that the Cartan-Maurer equations for the left invariant one-forms on
the quantum groups $U_q(N)$ can be written as:

$$dA^i+\PA{kl}{mn} \c{kl}{i} A^m \we A^n=0 $$

\noi so that the right-hand side of eq. (13) is to be expected
(see ref.[9]
for a discussion on quantum Cartan-Maurer equations and $q$-curvatures).
\sk

We next define the gauge variations:

$$\de \Am=-\dm \epsi - \Am \epsi + \epsi \Am \eqno(16)$$

\noi with

$$\epsi \equiv \epsi^i T_i $$

\noi and postulate the commutations

$$\epsi^i \Am^j = \R{ij}{kl} \Am^k \epsi^l  \eqno(17)$$

Under the variations (16) the field strength transforms as:

$$\de \Fmn = \epsi \Fmn - \Fmn \epsi   \eqno(18)$$

\noi Indeed the calculation is identical to the usual one.

Using (17) the gauge
variations of $\Am^i$ become

$$\de \Am^i = - \dm \epsi^i - \Am^j \epsi^k \C{jk}{i} \eqno(19)$$

Let us derive the commutations of $\epsi^i$ with $\Fmn^j$. To be as
general as possible we rewrite the $q$-commutations (17) as:

$$\epsi^i A^j = \R{ij}{kl} A^k \epsi^l  \eqno(20)$$

\noi i.e. we
do not use (6): our
results will be valid even for $q$-spacetime. We prove now that:

$$\epsi^i F^j = \R{ij}{kl} F^k \epsi^l  \eqno(21)$$

\noi Taking the
exterior derivative of (20) yields (we omit wedge symbols in the
following):

$$(d \epsi^i) A^j + \epsi^i d A^j = \R{ij}{kl}
   [(d A^k) \epsi^l + A^k (d \epsi^l)] \eqno(22) $$

\noi Requesting that the $d \epsi$ terms separately cancel
leads to:

$$(d \epsi^i)A^j=\R{ij}{kl} A^k (d \epsi^l) \eqno(23a)$$
$$\epsi^i d A^j = \R{ij}{kl} (d A^k)\epsi^l  \eqno(23b)$$

\noi Eq. (23b) tells us that the $d A^i$ part of $F^i$
does satisfy (21). What happens to the quadratic piece ? We should
find that

$$\epsi^i A^m A^n [\C{mn}{j}-
\Rinv{pq}{mn} \C{pq}{j}]=\R{ij}{kl} A^m A^n [\C{mn}{k}-
\Rinv{pq}{mn} \C{pq}{k}] \epsi^l  \eqno(24)$$

\noi in order to satisfy (21). The left-hand side
can be ordered as $AA\epsi$
by using (20) twice, and comparing it to the right-hand side we find
an equation of the type
$\La\La \Cb - \La \La \La^{-1} \Cb=\La \Cb -  \La \La^{-1} \Cb$. The
first and the third term cancel
because of the bicovariance condition (2d); the
second and the fourth term give:

$$\R{ir}{mk} \R{ks}{nl} \Rinv{pq}{rs}=\R{ij}{kl}
\Rinv{pq}{mn}\C{pq}{k} \eqno(25)$$

\noi Multiplying both sides
by $\R{mn}{uv}$ and using the Yang-Baxter equation
(2b) leads again to the bicovariance
condition (2d). Thus the $\epsi,F$ commutations (21)
hold because of the bicovariance conditions (2b) and (2d), and
can be used in eq. (18) to yield:

$$ \de \Fmn^i=-\Fmn^j \epsi^k \C{jk}{i}$$

\sk
Does the condition (2c) also play a
role ? The answer is yes: it ensures that the
commutation relations (4)
(or equivalently (5)) are preserved under the $q$-gauge
transformations (16). Indeed the
gauge variation of (5) is:

$$\de A^i A^j+ A^i \de A^j =
-{1\over{q^2+q^{-2}}} (\La+\La^{-1})^{ij}_{~~kl}
(\de A^k A^l + A^k \de A^l)\eqno(26)$$

\noi Substitute the variations
given by (16) and consider first the terms
without derivatives,
after having them
reordered as $AA\epsi$ via eq. (20). Using (5) on the left-hand side
leads exactly to the
bicovariance condition (2c).  The terms in (26) with the $d\epsi$
derivatives can be reordered as $A(d\epsi)$ using (23a), and
are easily seen to cancel
because of the cubic relation (9).
\sk
The commutations between $F^i$ and $A^j$ can be obtained by taking
the exterior derivative of the $A,A$ commutations (5):

$$F^i  A^j-A^i  F^j = -{1\over{q^2+q^{-2}}}
(\La+\La^{-1})^{ij}_{~~kl}
(F^k  A^l-A^k  F^l)\eqno(27)$$

Indeed the terms trilinear in $A$ that arise after using $dA=R-CAA$
do cancel, since they cancel in the case $F^i=0$, i.e. for left-
invariant $A^i$, and the wedge
products are unaltered in the case $F^i \not= 0$ (For a discussion
on the ``softening" of quantum group geometry, see ref. [9]).
\sk
Finally, we construct the $q$-lagrangian invariant
under the $U_q(N)$ quantum Lie algebra. We set:

$$ \Lcal = \Fmn^i \Fmn^j g_{ij}     \eqno(28)$$

\noi where $\gij$, the $q$-analogue
of the Killing metric, is determined by
requiring the invariance of $\Lcal$ under the $q$-gauge transformations
(16).
Under these transformations, the variation of the lagrangian (28) is
given by

$$\de \Lcal = -\C{mn}{i} \Fmn^m \epsi^n \Fmn^j \gij - \C{mn}{j}
\Fmn^i \Fmn^m \epsi^n \gij \eqno(29) $$

\noi After using (21) for reordering the terms as $FF\epsi$ we find that
$\de \Lcal$ vanishes when

$$\C{mn}{i} \R{nj}{rs} \gij + \C{rs}{j} g_{mj} = 0 \eqno(30)$$

We have not yet succeded in finding the general solution
for the $q$-Killing metric $\gij$. However, eq. (30) is
not difficult to solve in particular cases. For example,
in the $U_q(2) \equiv [SU(2) \times U(1)]_q$ case of
ref. [3], we find that

$$\gij =
   \left( \matrix{
     -2(1-q^2) + 2r & -2 & 0 & 0 \cr
     -2             & 2r & 0 & 0 \cr
      0             &  0 & 0 & 2q^2 (r+1) \cr
      0             &  0 & 2q^2 (r+1) & 0 \cr
   }\right)
\eqno(31)$$

\noi is the most general q-metric satisfying (30), depending on the
deformation parameter $q$ and on an extra arbitrary parameter $r$,
related
to the ratio of the $SU(2)$ and $U(1)$ coupling constants
(see ref. [3]).
For $q\rightarrow 1,~r\rightarrow 1$ one recovers the usual Killing
metric of \su .
\sk
{\sl Note 1:} It is clear that
$\gij$ should generalize the $q=1$ expression for
the Killing metric $\gij=Tr(\Cb_i \Cb_j)= \C{ri}{s} \C{sj}{r}$.
A quantum trace in the adjoint representation is needed.
\sk
{\sl Note 2:} The results of this Letter could be extended to the
quantum
groups of the $B,C,D,$ series.
\vfill\eject

{\bf References}
\sk
\item{[1]} A.P. Isaev and Z. Popowicz, {\sl q-Trace for the Quantum
Groups and q-deformed Yang-Mills Theory}, Wroclaw preprint ITP UWr
786-91; I.Y. Aref'eva and I.V. Volovich, Mod. Phys. Lett {\bf A6}
(1991) 893; Phys. Lett. {\bf B264} (1991) 62; S. Meljanac and
S. Pallua, {\sl Classical field lagrangian and deformed algebras},
DFPF/92/TH/15; A. Dimakis and F. M\"uller-Hoissen, {\sl Noncommutative
differential calculus, gauge theory and gravitation}, GOET-TP-33/92;
K. Wu and R.-J. Zhang, Commun. Theor. Phys. {\bf 17} (1992) 175;
T. Brzezi\'nski and S. Majid, {\sl Quantum group gauge theory on quantum
spaces}, DAMTP/92-27 and {\sl Quantum group gauge theory on classical
spaces}, DAMTP/92-51.
\sk
\item{[2a]} V. Drinfeld, Sov. Math. Dokl. {\bf 32} (1985) 254.

\item{[2b]} M. Jimbo, Lett. Math. Phys. {\bf 10} (1985) 63; {\bf 11}
        (1986) 247.

\item{[2c]} L.D. Faddeev,
N.Yu. Reshetikhin and L.A. Takhtajan, Algebra and
Analysis, {\bf 1} (1987) 178.

\item{[2d]} S. Majid, Int. J. Mod. Phys. {\bf A5} (1990) 1.
\sk
\item{[3]} L. Castellani, Phys. Lett. {\bf B292} (1992) 93.
\sk
\item{[4]} S.L. Woronowicz, Publ. RIMS, Kyoto Univ., Vol. {\bf 23}, 1
(1987) 117; Commun. Math. Phys. {\bf 111} (1987) 613 and
Commun. Math. Phys. {\bf 122}, (1989) 125.
\sk
\item{[5]} D. Bernard, {\sl Quantum Lie
algebras and differential calculus on quantum groups}, Proc. 1990 Yukawa
Int. Seminar, Kyoto; Phys. Lett. {\bf 260B} (1991) 389.
\sk
\item{[6]} B. Jur\v{c}o, Lett. Math. Phys. {\bf 22} (1991) 177.
\sk
\item{[7]} U. Carow-Watamura, M. Schlieker, S. Watamura
and W. Weich, Commun. Math. Phys. {\bf 142} (1991) 605.
\sk
\item{[8]} B. Zumino,
{\sl Introduction to the Differential Geometry
of Quantum Groups}, LBL-31432 and UCB-PTH-62/91, notes
of a plenary talk given at the 10-th IAMP Conf., Leipzig (1991);
F. M\"uller-Hoissen, J. Phys. A {\bf 25} (1992) 1703;
X.D. Sun and S.K. Wang, {\sl Bicovariant
differential calculus
on quantum group $GL_q(n)$}, Worldlab-Beijing preprint
CCAST-92-04, ASIAM-92-07,
ASITP-92-12 (1992); A. Sudbery, Phys. Lett. {\bf B284} (1992) 61;
P. Schupp, P. Watts and B. Zumino, {\sl Differential
Geometry on Linear Quantum Groups}, preprint LBL-32314, UCB-PTH-92/13
(1992) and {\sl Bicovariant quantum algebras and quantum Lie algebras},
LBL-32315, UCB-PTH-92/14; B. Zumino, {\sl Differential calculus on
quantum spaces and quantum groups}, LBL-33249, UCB-PTH-92/41.
\sk
\item{[9]} P. Aschieri and
L. Castellani, {\sl An
introduction to non-commutative
differential geometry on quantum groups}, preprint
CERN-TH.6565/92, DFTT-22/92 (1992), to be publ. in Int. Jou.
Mod. Phys. A.

\vfill\eject\end